\documentclass[prd,aps,floats,preprint,preprintnumbers]{revtex4}
\usepackage{amsmath,amssymb}
\usepackage{graphicx}
\usepackage{graphics}
\usepackage{epsfig}
\def\beq{\begin{equation}}
\def\eeq{\end{equation}}
\def\bea{\begin{eqnarray}}
\def\eea{\end{eqnarray}}
\def\nn{\nonumber} 
\begin{document}
\preprint{SU-4252-855 \vspace{1cm}}
\setlength{\unitlength}{1mm}
\title{Quantum Fields on the Groenewold-Moyal Plane: {\bf C}, {\bf P}, {\bf T} and {\bf CPT} \vspace{0.5cm}}
\author{E. Akofor$^{a}$}\thanks{eakofor@phy.syr.edu}\author{ A. P.
Balachandran$^{a}$}\thanks{bal@phy.syr.edu} \author{S. G.
Jo$^{a,b}$}\thanks{sgjo@knu.ac.kr} \author{A.
Joseph$^{a}$}\thanks{ajoseph@phy.syr.edu}
\affiliation{$^{a}$Department of Physics, Syracuse University,
Syracuse, NY 13244-1130, USA\\
$^{b}$Department of Physics, Kyungpook National University, Daegu, 702-701,
Korea\vspace{1cm}}\thanks{Permanent address}
\begin{abstract}
\vspace{0.1cm}
We show that despite the inherent non-locality of quantum field theories on the Groenewold-Moyal (GM) plane, one can find a class of ${\bf C}$, ${\bf P}$, ${\bf T}$ and ${\bf CPT}$ invariant theories. In particular, these are theories without gauge fields or with just gauge fields and no matter fields. We also show that in the presence of gauge fields, one can have a field theory where the Hamiltonian is ${\bf C}$ and ${\bf T}$ invariant while the $S$-matrix violates ${\bf P}$ and ${\bf CPT}$.

In non-abelian gauge theories with matter fields such as the electro-weak and $QCD$
sectors of the standard model of particle physics, ${\bf C}$, ${\bf P}$, ${\bf T}$ and the product of any pair of them  are broken while ${\bf CPT}$ remains intact for the case $\theta^{0i} =0$. (Here $x^{\mu} \star x^{\nu} - x^{\nu} \star x^{\mu} = i \theta^{\mu \nu}$, $x^{\mu}$: coordinate functions, $\theta^{\mu \nu} = -\theta^{\nu \mu}=$ constant.) When $\theta^{0i} \neq 0$, it contributes to breaking also ${\bf P}$ and ${\bf CPT}$. It is known that the $S$-matrix in a non-abelian theory depends on $\theta^{\mu \nu}$ only through $\theta^{0i}$. The $S$-matrix is frame dependent. It breaks (the identity component of the) Lorentz group. All the noncommutative effects vanish if the scattering takes place in the center-of-mass frame, or any frame where $\theta^{0i}P^{\textrm{in}}_{i} = 0$, but not otherwise. ${\bf P}$ and ${\bf CPT}$ are good symmetries of the theory in this special case. 
\end{abstract}
\maketitle
\section{INTRODUCTION}\label{sec:intro}
The Groenewold-Moyal plane or GM plane ${\cal {A}}_{\theta}(\mathbb{R}^{N})$ is the algebra of smooth
functions on $\mathbb{R}^{N}$ with the $\star$-product
\beq
\label{eq:starprdt}
f \star g = f e^{{i \over 2} \overleftarrow{\partial}_{\mu} \theta^{\mu \nu}
\overrightarrow{\partial}_{\nu}} g,
\eeq
\beq
\theta^{\mu \nu} = - \theta^{\nu \mu} = \text{constant.}\nn
\eeq

If $x=(x^{0}, x^{1}, ..., x^{N-1})$ labels a point on $\mathbb{R}^{N}$ and
$\hat{x}^{\mu}$ are coordinate functions, 
\beq
\label{eq:coordinate}
\hat{x}^{\mu}(x)=x^{\mu},
\eeq
eqn.~(\ref{eq:starprdt}) implies the commutation relation
\beq
\label{eq:comm-rel}
(\hat{x}^{\mu} \star \hat{x}^{\nu} - \hat{x}^{\nu} \star
\hat{x}^{\mu})=[\hat{x}^{\mu}, \hat{x}^{\nu}]_{\star}=i \theta^{\mu \nu}.
\eeq

Following Drinfel'd's original work \cite{drinfeld}, Chaichian et al.
\cite{chaichian} and Aschieri
et al. \cite{aschieri}, have shown that the diffeomorphism group ${\cal
D}(\mathbb{R}^N)$ of $\mathbb{R}^N$ acts on ${\cal A}_\theta(\mathbb{R}^N)$ provided
its coproduct $\Delta_\theta$ is the following twisted one:
\beq
\Delta_\theta(g) = {\cal F}^{-1}_{\theta}(g \otimes g){\cal
F}_{\theta},
\eeq
\beq
\label{eq:calF}
{\cal F}_{\theta} = e^{{i \over 2} \partial_{\mu} \otimes \theta^{\mu\nu}
\partial_{\nu}}.
\eeq
where $g$ is the group element and ${\cal F}_{\theta}$ is called the Drienfel'd twist \cite{drinfeld}.

The Poincar\'e group ${\cal P}$ is a subgroup of ${\cal D}(\mathbb{R}^N)$. Previous papers \cite{dimitrijevic, chaichian2, oeckl, bal, uv-ir}, examined quantum field theories (qft's) on the GM plane  ${\cal A}_\theta(\mathbb{R}^N)$ which are deformations of the Poincar\'e invariant qft's for $\theta^{\mu\nu}=0$.  It focused on the identity component ${\cal P}_{+}^{\uparrow}$ of ${\cal P}$. Using the twisted coproduct for ${\cal P}_{+}^{\uparrow}$, the following was proved \cite{chaichian, aschieri,
bal-stat, bal-sasha-babar} in a particular approach to gauge theories: $i.)$ in the absence of gauge fields, these
theories are Poincar\'e invariant. $ii.)$ Poincar\'e invariance is maintained also by
abelian gauge theories (with or without matter) and non-abelian gauge theories
without matter. $iii.)$  Poincar\'e invariance is lost in non-abelian gauge theories
with matter if $\theta^{0i}\neq 0$ while it is maintained if $\theta^{0i} = 0$.

Parity ${\bf P}$, time reversal ${\bf T}$ and ${\bf PT}$ are not elements of ${\cal P}_{+}^{\uparrow}$ \cite{pct}. We extend the previous analysis to  ${\bf P}$, ${\bf T}$ and ${\bf PT}$
as well here. The extension proves to be trivial in the absence of gauge fields. With gauge
fields present, further analysis is needed especially as ${\bf T}$ is anti-unitary. We
show in this paper that if for $\theta^{\mu \nu} = 0$, ${\bf P}$ and ${\bf T}$ are good symmetries in non-gauge theories, abelian gauge theories with or without matter fields, and non-abelian gauge theories without matter fields, then they continue to be so for $\theta^{\mu \nu} \neq 0$. But in non-abelian gauge theories with matter fields, such as the standard model, ${\bf P}$ and ${\bf CPT}$ are necessarily broken in scattering processes.

The behaviour of qft's on  ${\cal A}_\theta(\mathbb{R}^N)$ under charge conjugation is not affected by $\theta^{\mu\nu}$. Thus ${\bf CPT}$ invariance is maintained in the qft's on ${\cal A}_\theta(\mathbb{R}^N)$ in non-gauge theories, abelian gauge theories with or without matter fields and non-abelian gauge theories without matter fields. That is so even though they violate many of the axioms of local quantum field theories.

Discrete transformations for qft's on noncommutative spacetimes have been analysed previously by Sheikh-Jabbari and by \'Alvarez-Gaum\'e and V\'azquez-Mozo \cite{jabbari, gaume-mozo}. The qft's they analysed are however formulated differently from the ones we study here.
\section{A primer of past work.}
In this section, we summarize the pertinent aspects of our previous work for matter and gauge fields. While our treatment of matter fields is fully coherent with the work of Aschieri et al. \cite{aschieri}, the two treatments differ in the treatment of gauge fields. 
\subsection{Matter fields without gauge interactions}
We focus on a free complex scalar quantum field $\varphi$ as an example. The discussion can be adapted to any free matter field.

The field $\varphi$ on the GM plane ${\cal A}_\theta(\mathbb{R}^N)$ has the expansion
\beq
\varphi_{\theta} = \int d\mu(p)\; (a_{{\bf p}} \; e_{p} + b^{\dagger}_{{\bf p}}\; e_{-p}),\nn
\eeq
\beq
d\mu(p) = {d^{N-1}p \over 2p_{0}}, \; e_{p}(x)=e^{-ipx}, \;  p_{0}=\sqrt{{\bf
p}^{2}+m^{2}},\; \; \; m= \text{mass of $\varphi$}.
\eeq

We set $N=4$ for specificity. The operators $a_{{\bf p}}$, $b_{{\bf p}}$ can be
written in terms
of the annihiliation-creation operators $c_{{\bf p}}$, $d_{{\bf p}}$ for
$\theta^{\mu \nu}=0$ as follows using the ``dressing transformation" \cite{Grosse, Faddeev-Zamolodchikov}:
\beq
a_{{\bf p}} = c_{{\bf p}} \; e^{-{i \over 2}p_{\mu} \theta^{\mu\nu}P_{\nu}},\; \; \;
b_{{\bf p}} =
d_{{\bf p}}\; e^{-{i \over 2}p_{\mu}\theta^{\mu\nu}P_{\nu}},\nn
\eeq
where
\beq
P_{\mu} = \int {d^{3}p \over 2p_{0}}\; (c_{{\bf p}}^\dagger c_{{\bf p}} +
d^{\dagger}_{{\bf p}}
d_{{\bf p}})\; p_{\mu} = \text{Four-momentum operator}.\nn
\eeq

The commutation relations of $c_{{\bf p}}$, $c^{\dagger}_{{\bf p}}$, $d_{{\bf p}}$,
$d_{{\bf p}}^{\dagger}$ are standard,
\beq
[c_{{\bf p}},c_{{\bf q}}^{\dagger}]=[d_{{\bf p}},d_{{\bf q}}^{\dagger}]=2p_{0} \;
\delta^{3}({\bf p}-{\bf q}).
\eeq

The remaining commutators involving these operators vanish. 

The new operators $a_{{\bf p}}$ and $b_{{\bf p}}$ are called the twisted or dressed operators and the map from $c$, $d$- to the $a$, $b$- operators is called dressing transformation. (The Grosse-Faddeev-Zamolodchikov algebra is a generalization of the above twisted or dressed algebra \cite{Grosse, Faddeev-Zamolodchikov}. See also \cite{queiroz} in this connection.)

Note that $P_{\mu}$ can also be written in terms of the twisted operators:
\beq
\label{eq:pmu}
P_{\mu} =\int {d^{3}p \over 2p_{0}}\; (a_{{\bf p}}^{\dagger} a_{{\bf p}} +
b^{\dagger}_{{\bf p}}
b_{{\bf p}}) \; p_{\mu} = \text{Four-momentum}.
\eeq
That is because $p_{\mu} \theta^{\mu\nu}P_{\nu}$ commutes with any of the operators for momentum $p$. For example $[P_{\mu},a_{{\bf p}}]=-p_{\mu} a_{{\bf
p}}$ so that $[p_{\nu} \theta^{\nu\mu}
P_{\mu}, a_{{\bf p}}]= p_{\nu}\theta^{\nu\mu}p_{\mu}=0$, $\theta$ being antisymmetric. 

The antisymmetry of $\theta^{\mu \nu}$ allows us to write
\beq
c_{\bf p}e^{-{i \over 2}p_{\mu} \theta^{\mu \nu} P_{\nu}} = e^{-{i \over 2}p_{\mu}
\theta^{\mu \nu} P_{\nu}} c_{\bf p}, 
\eeq
\beq
c^{\dagger}_{\bf p}e^{{i \over 2}p_{\mu} \theta^{\mu \nu} P_{\nu}} = e^{{i \over
2}p_{\mu} \theta^{\mu \nu} P_{\nu}}c^{\dagger}_{\bf p}. 
\eeq

Hence the ordering of factors here is immeterial. 

It should also be noted that the map from $c$- to the $a$-operators is invertible,
\beq
c_{\bf p} = a_{\bf p} \; e^{{i \over 2}p_{\mu} \theta^{\mu\nu}P_{\nu}},\; \; \;
d_{\bf p} =
b_{\bf p}\; e^{{i \over 2}p_{\mu}\theta^{\mu\nu}P_{\nu}},\nn
\eeq
where $P_{\mu}$ is written as in eqn.~(\ref{eq:pmu}). 

The $\star$-product between the fields is
\beq
(\varphi_{\theta} \star \varphi_{\theta})(x) = \varphi_{\theta}(x) e^{{i \over 2} \overleftarrow{\partial}
\wedge \overrightarrow{\partial}} \varphi_{\theta}(y) |_{x=y},
\eeq
\beq
\overleftarrow{\partial} \wedge \overrightarrow{\partial} :=
\overleftarrow{\partial}_{\mu} \theta^{\mu \nu} \overrightarrow{\partial}_{\nu}.\nn
\eeq

The twisted quantum field $\varphi$ differs from the untwisted quantum field
$\varphi_{0}$ in two ways: $i.)$ $e_{p} \in {\cal A}_{\theta}(\mathbb{R}^{4})$ and
$ii.)$ $a_{{\bf p}}$ is twisted by statistics.

Both can be accounted by writing \cite{bal-sasha-babar}
\beq
\varphi_{\theta} = \varphi_{0} \; e^{{1 \over 2} \overleftarrow{\partial} \wedge P},
\label{eq:gaugefield}
\eeq
where $P_{\mu}$ is the total momentum operator. From this follows that the
$\star$-product of an
arbitrary number of fields $\varphi_{\theta}^{(i)}$ ($i$ = 1, 2, 3, $\cdots$) is
\beq
\varphi_{\theta}^{(1)} \star \varphi_{\theta}^{(2)} \star {\cdots} =(\varphi^{(1)}_{0}\varphi^{(2)}_{0} {\cdots}) \; e^{{1
\over 2} \overleftarrow{\partial} \wedge P}.
\label{eq:productfields}
\eeq

Although the rule (\ref{eq:gaugefield}) is for a spin-zero massive scalar field, we can apply 
it to all bosonic and fermionic matter fields with any spin. This is very
convenient when we write the interaction Hamiltonian involving matter fields.

We can write the interaction Hamiltonian density for a pure matter field as
\beq
{\cal H}_{I \theta} = {\cal H}_{I 0} \; e^{{1 \over 2} \overleftarrow{\partial}
\wedge P}
\eeq
on using eqn.~(\ref{eq:productfields}). Thus statistics untwists the $\star$ in
${\cal H}_{I \theta}$. This is what leads to the
$\theta$-independence of $S$-operator in the absence of gauge fields \cite{bal, uv-ir, bal-stat, bal-sasha-babar}. Interaction terms involving matter fields are always in this form.
\subsection{Matter fields with gauge interactions}
This section is based on \cite{bal-sasha-babar}.

We assume that the gauge (and gravity) fields are associated with the ``commutative
manifold" ${\cal A}_{0}(\mathbb{R}^{4})$ whereas for Aschieri et al. \cite{aschieri} they are
associated with ${\cal A}_{\theta}(\mathbb{R}^{4})$. Matter
fields on ${\cal A}_{\theta}(\mathbb{R}^{4})$ must be transported by the connection
compatibly with  eqn.~(\ref{eq:gaugefield}), so a natural choice for the covariant
derivative is \cite{bal-sasha-babar}
\beq
D_{\mu} \varphi_{\theta} = (D_{\mu}^{c} \varphi_{0}) \; e^{{1 \over 2}
\overleftarrow{\partial} \wedge P},
\label{eq:covariant}
\eeq
where
\beq
D_{\mu}^{c} \varphi_{0} = \partial_{\mu}\varphi_{0} + A_{\mu}\varphi_{0}\; ,
\eeq
$P_{\mu}$ is the total momentum operator for all the fields and
\beq
A_{\mu}\varphi_{0}(x)=A_{\mu}(x)\varphi_{0}(x) \; \; \text{[point-wise multiplication]}.
\eeq

This is indeed the correct choice of $D_{\mu}$ as it preserves the statistics,
Poincar\'e and gauge invariance, and the requirement that $D_{\mu}$ is associated
with  ${\cal A}_{0}(\mathbb{R}^{N})$:
\begin{eqnarray}
[D_{\mu}, D_{\nu}] \varphi_{\theta} &=& \Big([D^{c}_{\mu}, D^{c}_{\nu}]\varphi_{0}\Big)e^{{1
\over 2}\overleftarrow{\partial} \wedge P} \\
&=&\Big(F_{\mu \nu}^{c}\varphi_{0}\Big)e^{{1 \over 2}\overleftarrow{\partial} \wedge
P}.
\end{eqnarray}
As $F_{\mu \nu}^{c}$ is the standard $\theta^{\mu \nu}=0$ curvature, our gauge field
is associated with ${\cal A}_{0}(\mathbb{R}^{N})$. (For Aschieri et al. \cite{aschieri} the
curvature would be the $\star$-commutator of $D_{\mu}$'s.) The gauge theory formulation we adopt here is fully explained in \cite{bal-sasha-babar}. It differs from the formulation of Aschieri et al. \cite{aschieri} (where covariant derivative is defined using star product) and has the advantage of being able to accomodate any gauge group and not just $U(N)$ gauge groups and their direct products. The gauge theory formulation we adopt here avoids multiplicity of fields that the expression for covariant derivates with $\star$ product entails.

In the single-particle sector (obtained by taking the matrix element of eqn.~(\ref{eq:covariant}) between vacuum and one-particle states), the $P$ term can be dropped and we get for a single particle wave function $f$ of a particle associated with $\varphi_{\theta}$,
\beq
D_{\mu}f(x) = \partial_{\mu}f(x)+A_{\mu}(x)f(x).
\eeq

Note that we can also write $D_{\mu}\varphi_{\theta}$ using $\star$-product: 
\beq
D_{\mu}\varphi_{\theta} = \Big(D_{\mu}^{c} e^{{1 \over 2} \overleftarrow{\partial} \wedge
P}\Big)\star \Big(\varphi_{0}e^{{1 \over 2} \overleftarrow{\partial} \wedge
P}\Big).
\eeq

Our choice of covariant derivative allows us to write the interaction Hamiltonian density 
for pure gauge fields as follows:
\beq
{\cal H}_{I \theta}^{^G} = {\cal H}_{I 0}^{^G}.
\eeq

For a theory with matter and gauge fields, the interaction Hamiltonian density splits into
\beq
{\cal H}_{I \theta} = {\cal H}^{^{M, G}}_{I \theta}+{\cal H}^{^G}_{I \theta},
\eeq

where
\bea
{\cal H}^{^{M, G}}_{I \theta}&=&{\cal H}^{^{M, G}}_{I 0} \; e^{{1 \over 2}
\overleftarrow{\partial} \wedge P},\nn \\
{\cal H}^{^G}_{I \theta}&=&{\cal H}^{^G}_{I 0}.
\eea

The matter-gauge field couplings are also included in ${\cal H}^{^{M, G}}_{I \theta}$.

In quantum electrodynamics ($QED$), ${\cal H}^{^G}_{I \theta}=0$. Thus the
$S$-matrix for the twisted $QED$ is the same for the untwisted $QED$:
\beq
S^{^{QED}}_{\theta}=S^{^{QED}}_{0}.
\eeq

In a non-abelian gauge theory, ${\cal H}^{^G}_{\theta}={\cal H}^{^G}_{0} \neq 0$, so that in the presence of nonsinglet matter fields \cite{bal-sasha-babar}. 
\beq
S^{^{M, G}}_{\theta} \neq S^{^{M, G}}_{0}.
\eeq

\section{On ${\bf C}$, ${\bf P}$, ${\bf T}$ and ${\bf CPT}$}
In this section we investigate ${\bf C}$, ${\bf P}$, ${\bf T}$ and ${\bf CPT}$ for ${\cal
A}_{\theta}(\mathbb{R}^{N})$ . The ${\bf CPT}$ theorem \cite{cpt} is very fundamental in
nature and all local relativistic quantum field theories are ${\bf CPT}$ invariant. Qft's
on the GM plane are non-local and so it is important to investigate the validity of the ${\bf CPT}$
theorem in these theories.

Here we first recall a fundamental result of earlier work \cite{bal-sasha-babar} that ${\bf C}$ and the Poincar\'e group transform $c_{\bf k}$'s and $d_{\bf k}$'s and their adjoints as in the untwisted theories. The induced transformations on the fields automatically imply the twisted coproduct in the matter sector, and of course the untwisted coproduct for gauge fields. This simple rule is proved for ${\cal P}_{+}^{\uparrow}$ in \cite{bal-sasha-babar}. It then implies the same rule for the full group generated by ${\bf C}$ and ${\cal P}$ by the group properties of that group. (We always try to preserve such group properties.) This rule is repeatedly used
below.

The matrix $\theta^{\mu \nu}$ is a constant antisymmetric matrix. We emphasise that in the approach using the twisted coproduct for the Poincar\'e group, $\theta^{\mu \nu}$ is {\it not} transformed by Poincar\'e transformarions or in fact by any other symmetry: they are truly constants. Nevertheless Poincar\'e invariance and other symmetries can be certainly recovered for Lagrangians invariant under the twisted symmetry actions at the level of classical actions and Wightman functions \cite{drinfeld, aschieri, dimitrijevic, bal-stat}.

\subsection{Transformation of Quantum Fields Under ${\bf C}$, ${\bf P}$ and ${\bf T}$ }
\subsubsection{Charge conjugation ${\bf C}$}
The coproduct \cite{chaichian, aschieri} for the charge conjugation operator ${\bf C}$ in the twisted case is the
same as the coproduct for ${\bf C}$ in the untwisted case since the charge conjugation
operator commutes with $P_{\mu}$. So, we write
\beq
\Delta_{\theta}({\bf C}) = \Delta_{0}({\bf C}) = {\bf C} \otimes {\bf C}.
\eeq

Under charge conjugation,
\beq
c_{{\bf k}} \stackrel{{\bf C}}\longrightarrow d_{{\bf k}}, \; 
a_{{\bf k}} \stackrel{{\bf C}}\longrightarrow b_{{\bf k}}
\eeq
where $a_{{\bf k}}=c_{{\bf k}} \; e^{-{i \over 2} k \wedge P}$
and  $b_{{\bf k}}=d_{{\bf k}} \; e^{-{i \over 2} k \wedge P}$.

Products of quantum fields on ${\cal A}_{\theta}(\mathbb{R}^{N})$ transform in the
same way as they would on ${\cal A}_{0}(\mathbb{R}^{N})$ under the ${\bf C}$ operation. Thus we have
\beq
\varphi_{\theta} \stackrel{{\bf C}}\longrightarrow \varphi^{_{\bf C}}_{0}\; e^{{1 \over 2} \overleftarrow{\partial} \wedge P}, \; \; \varphi_{0}^{_{\bf C}} = {\bf C} \varphi_{0} {\bf C}^{-1}. 
\eeq
while the product of two such fields $\varphi$ and $\chi$ transforms according to
\bea
\varphi_{\theta} \star \chi_{\theta} &=&(\varphi_{0} \chi_{0})\; e^{{1 \over 2} \overleftarrow{\partial} \wedge P} \nn \\ &\stackrel{{\bf C}}\longrightarrow& ({\bf C} \varphi_{0}\chi_{0}{\bf C}^{-1})\; e^{{1 \over 2} \overleftarrow{\partial} \wedge P} \nn \\
&=&(\varphi_{0}^{{\bf C}} \chi_{0}^{{\bf C}})\; e^{{1 \over 2}\overleftarrow{\partial} \wedge P}.
\eea
\subsubsection{Parity ${\bf P}$}
Parity is a unitary operator on ${\cal A}_{0}(\mathbb{R}^{N})$. But parity transformations do not induce automorphisms of ${\cal A}_{\theta}(\mathbb{R}^{N})$ \cite{bal-unitary} if its coproduct is 
\beq
\Delta_{0}({\bf P})={\bf P} \otimes {\bf P}.
\eeq
That is the coproduct is not compatible with the $\star$-product. Hence the coproduct for parity is not the same as that for the $\theta^{\mu \nu}=0$ case. 

But the twisted coproduct $\Delta_{\theta}$, where
\beq
\Delta_\theta({\bf P}) = {\cal F}_{\theta}^{-1} \; \Delta_{0} ({\bf P}) \; {\cal
F}_{\theta},
\eeq
{\it is} compatible with the $\star$-product. So, for ${\bf P}$ as well, compatibility with the $\star$-product fixes the coproduct \cite{bal}.

Under parity,
\beq
c_{{\bf k}} \stackrel{{\bf P}}\longrightarrow c_{-{\bf k}}, \; \; \; \; d_{{\bf k}}
\stackrel{{\bf P}}\longrightarrow d_{-{\bf k}}
\label{eq:cdP}
\eeq
and hence
\beq
a_{{\bf k}} \stackrel{{\bf P}}\longrightarrow a_{-{\bf k}} \;
e^{i(k_{0}\theta^{0i}P_{i}-k_{i}\theta^{i0}P_{0})}, \; \; \; \; b_{{\bf k}}
\stackrel{{\bf P}}\longrightarrow b_{-{\bf k}} \;
e^{i(k_{0}\theta^{0i}P_{i}-k_{i}\theta^{i0}P_{0})}
\label{eq:abP}
\eeq
By an earlier remark \cite{bal-sasha-babar}, eqns. (\ref{eq:cdP}) and (\ref{eq:abP}) imply the transformation law for twisted scalar fields. A twisted complex scalar field $\varphi_{\theta}$ transforms under parity as follows,
\bea
\varphi_{\theta}&=& \varphi_{0} \; e^{{1 \over 2}\overleftarrow{\partial} \wedge P}\nn \\
&\stackrel{{\bf P}}\longrightarrow& {\bf P}\Big(\varphi_{0} \; e^{{1 \over 2}\overleftarrow{\partial} \wedge P}\Big){\bf P}^{-1}\nn\\
&=&\varphi_{0}^{_{\bf P}}\; e^{{1 \over 2} \overleftarrow{\partial} \wedge (P_{0}, -\overrightarrow{P})},
\eea
where $\varphi_{0}^{_{\bf P}} = {\bf P} \varphi_{0} {\bf P}^{-1}$ and $\overleftarrow{\partial} \wedge (P_{0}, -\overrightarrow{P}) := -\overleftarrow{\partial}_{0}
\theta^{0i} P_{i} -\overleftarrow{\partial}_{i} \theta^{ij}P_{j}+ \overleftarrow{\partial}_{i} \theta^{i0}P_{0}$.

The product of two such fields $\varphi_{\theta}$ and $\chi_{\theta}$ transforms according to
\bea
\varphi_{\theta} \star \chi_{\theta} &=&(\varphi_{0} \chi_{0})\; e^{{1 \over 2} \overleftarrow{\partial} \wedge P} \nn \\ &\stackrel{{\bf P}}\longrightarrow& (\varphi^{_{\bf P}}_{0} \chi^{_{\bf P}}_{0})\; e^{{1 \over 2} \overleftarrow{\partial} \wedge (P_{0}, -\overrightarrow{P})}
\eea

Thus fields transform under ${\bf P}$ with an extra factor $e^{-(\overleftarrow{\partial}_{0}\theta^{0i}P_{i} + \partial_{i}\theta^{ij}P_{j})} = e^{-\overleftarrow{\partial}_{\mu}\theta^{\mu j}P_{j}}$ when $\theta^{\mu \nu} \neq 0$.
\subsubsection{Time reversal ${\bf T}$}
Time reversal ${\bf T}$ is an anti-linear operator. Due to antilinearity, ${\bf T}$ induces automorphisms on ${\cal A}_{\theta}(\mathbb{R}^{N})$ for any $\theta^{\mu \nu}$ as well
\cite{bal-unitary}.

Under time reversal,
\beq
c_{{\bf k}} \stackrel{{\bf T}}\longrightarrow c_{-{\bf k}}, \; \; \; \; d_{{\bf k}}
\stackrel{{\bf T}}\longrightarrow d_{-{\bf k}}
\eeq
\beq
a_{{\bf k}} \stackrel{{\bf T}}\longrightarrow a_{-{\bf k}} \;
e^{-i(k_{i}\theta^{ij}P_{j})}, \; \; \; \; b_{{\bf k}} \stackrel{{\bf T}}\longrightarrow
b_{-{\bf k}} \; e^{-i(k_{i}\theta^{ij}P_{j})}.
\eeq

Compatibility with the $\star$-product fixes the coproduct for ${\bf T}$ to be
\beq
\Delta_\theta({\bf T}) = {\cal F}_{\theta}^{-1} \; \Delta_{0} ({\bf T}) \; {\cal F}_{\theta}.
\eeq

This coproduct is also required in order to maintain the group properties of ${\cal P}$,
the full Poincar\'e group.

A twisted complex scalar field $\varphi_{\theta}$ hence transforms under time reversal as
follows,
\bea
\varphi_{\theta}&=& \varphi_{0} \; e^{{1 \over 2}\overleftarrow{\partial} \wedge P}\nn \\
&\stackrel{{\bf T}}\longrightarrow& {\bf T}\Big(\varphi_{0} \; e^{{1 \over 2}\overleftarrow{\partial} \wedge P}\Big){\bf T}^{-1}\nn \\
&=&\varphi_{0}^{_{\bf T}}\; e^{{1 \over 2} \overleftarrow{\partial} \wedge (P_{0}, -\overrightarrow{P})},
\eea
while the product of two such fields $\varphi_{\theta}$ and $\chi_{\theta}$ transforms according to
\bea
\varphi_{\theta} \star \chi_{\theta} &=&(\varphi_{0} \chi_{0})\; e^{{1 \over 2} \overleftarrow{\partial} \wedge P} \nn \\ &\stackrel{{\bf T}}\longrightarrow& (\varphi^{_{\bf T}}_{0} \chi^{_{\bf T}}_{0})\; e^{{1 \over 2} \overleftarrow{\partial} \wedge (P_{0}, -\overrightarrow{P})}
\eea

Thus the time reversal operation as well induces an extra factor $e^{-\overleftarrow{\partial}_{i}\theta^{ij}P_{j}}$ in the transformation
property of fields when $\theta^{\mu \nu} \neq 0$.
\subsubsection{${\bf CPT}$}
When ${\bf CPT}$ is applied,

\beq
c_{{\bf k}} \stackrel{{\bf CPT}}\longrightarrow d_{{\bf k}}, \; \; \; \; d_{{\bf k}}
\stackrel{{\bf CPT}}\longrightarrow c_{{\bf k}}
\eeq
\beq
a_{{\bf k}} \stackrel{{\bf CPT}}\longrightarrow b_{{\bf k}}e^{i(k \wedge P)}, \; \; \; \; 
b_{{\bf k}} \stackrel{{\bf CPT}}\longrightarrow a_{{\bf k}}e^{i(k \wedge P)}.
\eeq

The coproduct for ${\bf CPT}$ is of course
\beq
\Delta_\theta({\bf CPT}) = {\cal F}_{\theta}^{-1} \; \Delta_{0} ({\bf CPT}) \; {\cal F}_{\theta}.
\eeq

A twisted complex scalar field $\varphi_{\theta}$ transforms under ${\bf CPT}$ as follows,
\bea
\varphi_{\theta}&=& \varphi_{0} \; e^{{1 \over 2}\overleftarrow{\partial} \wedge P}\nn \\
&\stackrel{{\bf CPT}}\longrightarrow& {\bf CPT}\Big(\varphi_{0} \; e^{{1 \over 2}\overleftarrow{\partial} \wedge P}\Big) ({\bf CPT})^{-1}\nn \\
&=& \varphi_{0}^{_{{\bf CPT}}}\; e^{{1 \over 2} \overleftarrow{\partial} \wedge P},
\eea
while the product of two such fields $\varphi_{\theta}$ and $\chi_{\theta}$ transforms according to
\bea
\varphi_{\theta} \star \chi_{\theta} &=&(\varphi_{0} \chi_{0})\; e^{{1 \over 2} \overleftarrow{\partial} \wedge P} \nn \\ &\stackrel{{\bf CPT}}\longrightarrow& (\varphi^{_{{\bf CPT}}}_{0} \chi^{_{{\bf CPT}}}_{0})\; e^{{1 \over 2} \overleftarrow{\partial} \wedge P}.
\eea
\section{{\bf CPT} In  Non-Abelian Gauge Theories}
The standard model, a non-abelian gauge theory, is ${\bf CPT}$ invariant, but it is not invariant under ${\bf C}$, ${\bf P}$, ${\bf T}$ or products of any two of them. So we focus on discussing just ${\bf CPT}$ for its $S$-matrix when $\theta^{\mu \nu} \neq 0$. We also cover quantum electrodynamics ($QED$) by brief remarks. The discussion here can be easily adapted to any other non-abelian gauge theory.
\subsection{Matter fields and their couplings to gauge fields}
The interaction representation $S$-matrix is

\beq 
{S}^{^{M, G}}_{\theta} = \text{T} \; exp \; \Big[{-i\int d^{4}x \; {\cal H}^{^{M, G}}_{I
\theta}(x)}\Big]
\eeq
where ${\cal H}^{^{M, G}}_{I \theta}$ is the interaction Hamiltonian density for matter
fields (including also matter-gauge field couplings). Under ${\bf CPT}$,
\beq
{\cal H}^{^{M, G}}_{I \theta}(x) \stackrel{{\bf CPT}}\longrightarrow {\cal H}^{^{M, G}}_{I
\theta}(-x)e^{\overleftarrow{\partial} \wedge P}
\eeq
where $\overleftarrow{\partial}$ has components ${\overleftarrow{\partial} \over \partial x_{\mu}}$. We write ${\cal H}^{^{M, G}}_{I \theta}$ as

\beq
{\cal H}^{^{M, G}}_{I \theta} = {\cal H}^{^{M, G}}_{I 0} \; e^{{1 \over 2}
\overleftarrow{\partial} \wedge P}.
\label{eq:matter}
\eeq
Thus we can write the interaction Hamiltonian density after ${\bf CPT}$ transformation in
terms of the untwisted interaction Hamiltonian density:

\bea
{\cal H}^{^{M, G}}_{I \theta}(x) \; \; \stackrel{{\bf CPT}}\longrightarrow&& {\cal H}^{^{M, G}}_{I
\theta}(-x)\; e^{\overleftarrow{\partial} \wedge P} \nn \\ &=& {\cal
H}^{^{M, G}}_{I0}(-x)\; e^{-{1 \over 2}\overleftarrow{\partial} \wedge
P}\; e^{\overleftarrow{\partial} \wedge P}\nn \\
&=&{\cal H}^{^{M, G}}_{I0}(-x)\; e^{{1 \over 2}\overleftarrow{\partial} \wedge P}
\eea

Hence under ${\bf CPT}$,

\beq
{S}^{^{M, G}}_{\theta} = \text{T} \; exp \; \Big[-i\int d^{4}x \; {\cal H}^{^{M, G}}_{I
0}(x) \; e^{{1 \over 2}\overleftarrow{\partial}\wedge P}\Big] \rightarrow \text{T} \; exp \; \Big[i\int d^{4}x \; {\cal H}^{^{M, G}}_{I
0}(x) \; e^{-{1 \over 2}\overleftarrow{\partial}\wedge P}\Big] = ({S}^{^{M, G}}_{-\theta})^{-1} \nn
\eeq

But it has been shown elsewhere that ${S}^{^{M, G}}_{\theta}$ is independent of
$\theta$ \cite{uv-ir}. Hence also ${S}^{^{M, G}}_{\theta}$ is independent of
$\theta$. 

Therefore a qft with no pure gauge interaction is ${\bf CPT}$ ``invariant" on ${\cal
A}_{\theta}(\mathbb{R}^{N})$. In particular quantum electrodynamics ($QED$)
preserves ${\bf CPT}$.

\subsection{Pure Gauge Fields}
The interaction Hamiltonian density for pure gauge fields is independent of
$\theta^{\mu \nu}$ in the approach of \cite{bal-sasha-babar}:
\beq
{\cal H}_{I \theta}^{^G} = {\cal H}_{I 0}^{^G}\; .
\eeq

Hence also the $S$-matrix becomes $\theta$-independent, 

\beq 
{S}^{^G}_{\theta} = {S}^{^G}_{0},
\eeq

and ${\bf CPT}$ holds as a good ``symmetry" of the theory.
\subsection{Matter and Gauge Fields}
All interactions of matter and gauge fields can be fully discussed by writing the $S$-matrix as\beq 
{{\bf S}}^{^{M,G}}_{\theta} = \text{T} \; exp \; \Big[{-i\int d^{4}x \; {\cal
H}_{I
\theta}(x)}\Big],
\eeq
\beq
{\cal H}_{I \theta} = {\cal H}^{^{M, G}}_{I \theta}+{\cal H}^{^G}_{I \theta},
\eeq
where
\beq
{\cal H}^{^{M, G}}_{I \theta}={\cal H}^{^{M, G}}_{I 0} \; e^{{1 \over 2}
\overleftarrow{\partial} \wedge P}\nn
\eeq
and
\beq
{\cal H}^{^G}_{I \theta}={\cal H}^{^G}_{I 0}\; .\nn
\eeq

In $QED$, ${\cal H}^{^G}_{I \theta}=0$. Thus the $S$-matrix ${\bf S}^{^{QED}}_{\theta}$ is
the
same as for the $\theta^{\mu \nu} =0$ case in $QED$:
\beq
{\bf S}^{^{QED}}_{\theta}={\bf S}^{^{QED}}_{0}.
\eeq

Hence ${\bf C}$, ${\bf P}$, ${\bf T}$ and ${\bf CPT}$ are good ``symmetries" for $QED$ on the GM plane.

For a non-abelian gauge theory with non-singlet matter fields, ${\cal H}^{^G}_{I \theta}={\cal H}^{^G}_{I 0} \neq 0$ so that if ${\bf S}^{^{M, G}}_{\theta}$ is the $S$-matrix of the theory,
\beq
{\bf S}^{^{M, G}}_{\theta} \neq {\bf S}^{^{M, G}}_{0}.
\eeq

The $S$-matrix ${\bf S}^{^{M,G}}_{\theta}$ depends only on $\theta^{0i}$ in a
non-abelian theory (see section V), that is,  ${\bf S}^{^{M,G}}_{\theta^{\mu \nu}} =
{\bf S}^{^{M,G}}_{\theta^{0i}}$. Applying ${\bf C}$, ${\bf P}$ and ${\bf T}$ on ${\bf S}^{^{M,G}}_{\theta}$ we can see that ${\bf C}$ and ${\bf T}$ do not affect $\theta^{0i}$ while ${\bf P}$ changes its sign. Thus a non-zero $\theta^{0i}$ contributes to ${\bf P}$ and ${\bf CPT}$ violation.
\subsection{On $QED$}
The $S$-matrix of $QED$ is invariant under ${\bf C}$, ${\bf P}$ and ${\bf T}$ for $\theta^{\mu \nu} = 0$. But as remarked earlier, its $S$-matrix is independent of $\theta^{\mu \nu}$.
Hence $QED$ is also ${\bf C}$, ${\bf P}$ and ${\bf T}$ invariant for any $\theta^{\mu \nu}$. 
\section{On Feynman Graphs}
The work of this section overlaps with \cite{bal-sasha-babar} and \cite{bal-sasha-queiroz} where Feynman rules are fully developed and field theories are analyzed further. 

In non-abelian gauge theories, ${\cal H}^{^{G}}_{I \theta}={\cal H}^{^{G}}_{I 0}$ is not zero as gauge fields have self-interactions. The preceding discussions show that the effects of $\theta^{\mu \nu}$ can show up only in Feynman diagrams which are sensitive to products of ${\cal H}^{^{M, G}}_{I \theta}$'s with ${\cal H}^{^{G}}_{I 0}$'s. Fig. 1 shows two such diagrams.
\begin{figure}
  \begin{center}
 \includegraphics[scale=1]{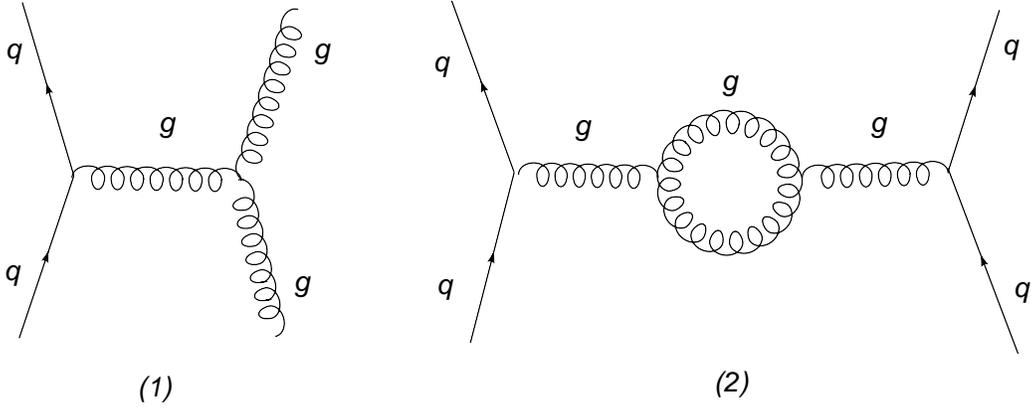}
  \end{center}
  \caption{${\bf CPT}$ violating processess on GM plane. (1) shows quark-gluon scattering with a three-gluon vertex. (2) shows a gluon-loop contribution to quark-quark scattering. (Calculation of such processes are being attempted).
  }
  \label{cptViolatingDiagram}
\end{figure} 

As an example, consider the first diagram in Fig. 1. To lowest order, it depends on $\theta^{0i}$. 

We can substitute eqn. (\ref{eq:matter}) for ${\cal H}^{^{M, G}}_{I \theta}$ and integrate over ${\bf x}$. That gives,
\beq
{\bf S}^{(2)} =-{1 \over 2} \int d^{4}x d^{4}y \;  \textrm{T} \Big(H_{I 0}^{^{M, G}}(x)\; e^{{1 \over 2} \overleftarrow{\partial}_{0} \theta^{0i} P_{i}}H_{I 0}^{^G}(y)\Big)\nn
\eeq
where $\overleftarrow{\partial}_{0}$ acts {\it only} on $H_{I 0}^{^{M, G}}(x)$ (and not on the step functions in time entering in the definition of $\textrm{T}$.)

Now $P_{i}$, being component of spatial momentum, commutes with 
\beq
\int d^{3}y \; {\cal H}^{^{G}}_{I 0}(y)\nn
\eeq
and hence for computing the matrix element defining the process ({\it 1}) in Fig. 1, we can substitute $\overrightarrow{P}_{\textrm{in}}$ for $\overrightarrow{P}$,  $\overrightarrow{P}_{\textrm{in}}$ being the total incident spatial momentum: 
\beq
{\bf S}^{(2)} =-{1 \over 2} \int d^{4}x d^{4}y \;  \textrm{T} \Big(H_{I 0}^{^{M, G}}(x)\; e^{{1 \over 2} \overleftarrow{\partial}_{0} \theta^{0i} P^{\textrm{in}}_{i}}H_{I 0}^{^G}(y)\Big).
\eeq

Thus ${\bf S}^{(2)}$ depends on $\theta^{0i}$ unless
\beq
\theta^{0i}P^{\textrm{in}}_{i} = 0.
\eeq

This will happen in the center-of-mass system or more generally if $\overrightarrow{\theta^{0}} = $($\theta^{01}$, $\theta^{02}$, $\theta^{03}$) is perpendicular to $\overrightarrow{P}^{\textrm{in}}$.

Under ${\bf P}$ and ${\bf CPT}$, $\theta^{0i} \rightarrow -\theta^{0i}$. This shows clearly that in a general frame, $\theta^{0i}$ contributes to ${\bf P}$ violation and causes ${\bf CPT}$ violation. 

The dependence of $S^{(2)}$ on the incident total spatial momentum shows that 
the scattering matrix is not Lorentz invariant. This noninvariance 
is caused by the nonlocality of the interaction Hamiltonian density: if we 
evaluate it at two spacelike separated points, the resultant operators do 
not commute. Such a violation of causality 
can lead to Lorentz-noninvariant $S$-operators \cite{bal-sasha-babar}.

The reasoning which reduced $e^{{1 \over 2}\overleftarrow{\partial} \wedge P}$ to $e^{{1 \over 2}\overleftarrow{\partial}_{0} \theta^{0i} P^{\textrm{in}}_{i}}$ is valid to all such factors in an arbitrary order in the perturbation expansion of the $S$-matrix and for arbitrary processes, $\overrightarrow{P}^{\textrm{in}}$ being the total incident spatial momentum. As $\theta^{\mu \nu}$ occur only in such factors, this leads to an interesting conclusion: if the scattering happens in the
center-of-mass frame, or any frame where $\theta^{0i}P^{\textrm{in}}_{i} = 0$, then the
$\theta$-dependence goes away from the $S$-matrix. That is, $P$ and $CPT$ remain intact if $\theta^{0i}P^{\textrm{in}}_{i} = 0$. The theory becomes $P$ and $CPT$ violating in all other frames. 
\section{Experimental Signals}
Terms with products of ${\cal H}^{^{M, G}}_{I \theta}$ and ${\cal H}^{^G}_{I \theta}$ are
$\theta$-dependent and they violate ${\bf CPT}$. Electro-weak and $QCD$ processes will
acquire dependence on $\theta$. This is the case when a diagram involves products of
${\cal H}^{^{M, G}}_{I \theta}$ and ${\cal H}^{^G}_{I \theta}$. For example
quark-gluon and quark-quark scattering on the GM plane become $\theta$-dependent ${\bf CPT}$
violating processes (See Fig.1). This may be tested experimentally. Possibilities in this
direction are being explored.
\section{Conclusions}
We have examined the discrete symmetries  ${\bf C}$, ${\bf P}$, ${\bf T}$ and ${\bf CPT}$ for qft's on the GM plane; showing how they are modified by the twisted statistics of quantum fields.
Twisted statistics is required by Lorentz invariance. We have shown that the action
of these discrete symmetries on the $S$-matrix is independent of $\theta^{\mu \nu}$
in particular when matter and non-abelian gauge fields interacting with each other are not
present. However in the presence of such matter and gauge fields, the $S$-matrix
violates ${\bf P}$ and ${\bf CPT}$. (It violates also Lorentz invariance
\cite{bal-sasha-babar}.) We have also mentioned some processes in which the
$\theta$-dependence is apparent. If the scattering happens in the
center-of-mass frame, or any frame where $\theta^{0i}P^{\textrm{in}}_{i} = 0$, then the
$\theta$-dependance goes away from the $S$-matrix. ${\bf P}$ and ${\bf CPT}$ remain intact if $\theta^{0i}P^{\textrm{in}}_{i} = 0$. 
\section{ACKNOWLEDGEMENTS}
We gratefully acknowledge the discussions with A. Pinzul and B. A.  Qureshi. This
work was partially supported by the US Department of Energy under grant number DE-FG02-85ER40231.
\bibliographystyle{apsrmp}

\end{document}